\documentclass[a4paper, 12pt, times]{article}
\usepackage{graphicx}
\usepackage{epstopdf}
\usepackage{caption}
\usepackage{subcaption}
\usepackage{subfig}
\usepackage{amsmath}
\usepackage{amssymb}
\usepackage{natbib}
\usepackage{fancyhdr}
\usepackage{titlesec}
\usepackage[font=footnotesize]{caption}

\newfont{\tit}{ptmb at 14pt}      
\newfont{\auth}{ptmr at 12pt}
\newfont{\sect}{ptmb at 12pt}
\newfont{\subsect}{ptmb at 11pt}

\titleformat{\section}{\sect}{\thesection}{1em}{}
\titleformat{\subsection}{\subsect}{\thesubsection}{1em}{}

\newcommand{\acknowledgements}[0]{\vspace{16pt} \noindent {\sect Acknowledgements} \vspace{10pt}\\}
\newcommand{\references}[0]{\vspace{16pt} \noindent {\sect References} \vspace{10pt}\\}
\renewcommand{\abstract}[0]{\noindent {\sect Abstract} \vspace{5pt}\\}

\begin{document}

\begin{center}
{Dynamics of a buoyant plume in a linearly stratified environment using simultaneous PIV-PLIF measurements}\\  

\vspace{10pt}

{ \underline{Harish N. Mirajkar$^1$} and Sridhar Balasubramanian$^{1,2}$}\\ 

\vspace{10pt}


{$^1$Department of Mechanical Engineering, \\
	$^{1,2}$ IDP in Climate Studies, 
Indian Institute of Technology Bombay, India}\\
sridharb@iitb.ac.in
\end{center}

\abstract The presence of stratified layer in atmosphere and ocean leads to buoyant vertical motions, commonly referred to as plumes. It is important to study the mixing dynamics of a plume at a local scale in order to model their evolution and growth. Such a characterization requires measuring the velocity and density of the mixing fluids simultaneously. Here, we present the results of a buoyant plume propagating in a linearly stratified medium with a density difference of 0.5 \%, thus  yielding a buoyancy frequency of $N$=0.15 $\mbox{s}^{-1}$. To understand the plume behaviour, statistics such as centerline and axial velocities along varying downstream locations, turbulent kinetic energy, Reynolds stress, and buoyancy flux were measured. The centerline velocity was found to decrease with increase in height. The Reynolds stress and buoyancy flux profiles showed the presence of a unstable layer and the mixing associated within that layer.

\section{Introduction} 
Buoyant jets and plumes are commonly seen in both environmental and industrial applications. They occur whenever a source of buoyancy creates a rising motion of fluid upwards and away from the source. A buoyant jet is an initial-momentum dominated flow that has a non-zero density difference between medium and the surrounding ambient fluid, and at some distance becomes buoyancy-dominated. Some common occurrences of buoyant jet and plume are in hydrothermal vents/oil spills in ocean, rising of ash plumes from the volcanic eruptions, ocean overflows and wastewater discharge into ocean. A variety of factors may influence the behavior of buoyant jets and plumes in field applications, including  cross flows of winds, stratification of ambient medium, and boundary interactions. While numerous studies on characterization of jets interacting with homogeneous environment exists, only a few focus on plume in a linearly stratified environment. A buoyant plume that interacts with a stably stratified environment will behave differently compared to a uniform environment. In stratified environment, mixing near the source of the plume will cause it to become denser, where the density of the surrounding fluid linearly decreases upwards. Eventually, the buoyancy force acting on the plume will change sign and become negative. The momentum will eventually diminish, where the plume succumbs to a finite height and spreads out horizontally, causing the plume to become two-dimensional. 

Pioneer research on this topic was first done by \cite{Morton56} to measure the maximum height, $Z_m$, for the buoyant plume in the stratified flow using the following flow parameters: the local buoyancy flux ($B_0$), the local momentum flux ($M_0$), and the buoyancy frequency ($N$), defined as $N=[-(g/ \rho_b)(\partial \rho/ \partial z)]^{1/2}$, where $\rho_b$ is the bottom density of the stratified tank and $\partial\rho/\partial z $ \  is the vertical density gradient. Following the seminal work of Morton et al, other researcher have also measured the plume dynamics in the stratified environment [see for e.g.\cite{Bloomfield98}]. \cite{Webster01} developed to measure simultaneously the velocity and concentration field using digital particle image velocimetry (DPIV) and planar laser-induced fluorescence (PLIF) for a turbulent jet in the uniform medium to measure the mean velocity, turbulent stresses, mean concentration variance.  
\cite{Liu01} studied the horizontal jet discharging into a linear stratified medium. The results for the stratified case, obtained using DPIV, showed that the mean centerline velocity decreases much more rapidly than the unstratified case, where
Reynolds stress profiles never reached a self-similar state, indicating that stratification
changes both the overall turbulence characteristics and mixing.

Recently, \cite{Xu12} observed a horizontal dense jet injected into a lighter
stratified solution using combined particle image velocimetry and planar laser
induced fluorescence. They studied flow  structure and mixing dynamics of the dense jet in the lighter solution. In a
different study by the same group [\cite{Xu13}], they simultaneously obtained velocity and
density fields by adjusting both the flow rate and Richardson number, changing the
jet buoyancy. Their results showed asymmetric velocity and buoyancy profiles, where
the mixing effect was reduced by the stably stratified environment. Despite the numerous study on this topic,  the dynamics of a buoyant plume in the linearly stratified medium has not been well studied at local scale. Therefore, the focus of our work to understand the local mixing dynamics of a buoyant plume using simultaneous PIV-PLIF technique. As a first cut, we only report the results at low buoyancy frequency.
\vspace{8pt} 

\vspace{8pt}

\section{Experimental set-up}
The experiments were carried out in a tank facility, whose configuration is illustrated in Fig. 1. The tank T2 is made of plexiglas, measuring 91 cm long by 91 cm wide by 60 cm high. The second tank (T1), a 60 cm cubical tank, was used as the reservoir for storing the jet fluid. The density of the jet fluid, $\rho_j$= 996.9 $kg/m^{3}$  was kept constant in all the experiments. The tank (T2) was linearly stratified using the double bucket technique as discussed in \cite{Oster63,Mirajkar15}.  The strength of the stratification was maintained to be 0.15$\mbox{s}^{-1}$. A portable densitometer ($Anton paar DMA 35$) was used to check the density in the two buckets. Density profile in the stratified tank was checked by collecting the samples at every intervals in the experimental tank to ensure the density profile is linear. 

\begin{figure}
\begin{center}
\includegraphics[scale=0.55,trim=15mm 0mm 0mm 0mm]{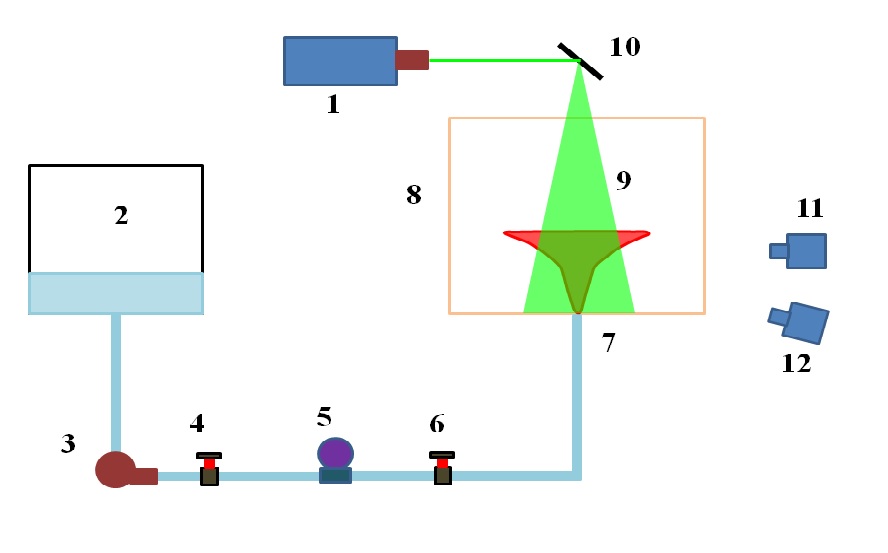}
\caption{Schematic of experimental setup. 1. Laser source, 2.Tank (T1), 3. Centrifugal pump, 4,6.Control valve, 7. Jet nozzle, 8.Experimental tank (T2), 9. Laser sheet, 10. Mirror, 11. PIV camera, 12. PLIF camera.}
\label{fig:1}
\end{center}
\end{figure}

A centrifugal pump was used to discharge the jet fluid into the ambient linearly stratified environment using a round jet nozzle fixed at the bottom of tank (T2). The jet nozzle was 160 $mm$ in length with diameter $D$ = 12.7 $mm$. It was made of aluminum and comprised of a diffuser, settling chamber, and a contraction section (\cite{Mehta1979}). A honeycomb was placed in the settling chamber to reduce the flow fluctuations and to generate a stable flow at the nozzle exit. The exit vertical velocity at the nozzle was maintained constant at $U_o$ = 17$cm/s$, thereby giving a jet Reynolds number $Re$= 2400, and the initial Richardsons number was, $Ri_{b}=(\frac{\pi}{4})^{0.25}\sqrt{\frac{g'D}{U_o^2}}$=0.058. A linear stable stratification with salt-water-ethanol mixture was obtained in T2, such that heavy fluid settles at the bottom and lighter fluid on the top. Once the fluid is filled into tank T2 using the two-bucket technique, it is allowed to stabilize for approximately 2 hours to achieve stable uniform linear stratification with height. A combination of Particle Image Velocimetry (PIV) and Planar Laser Induced Fluorescence (PLIF) is used for simultaneous velocity and density measurements as illustrated in figure 1. Before the experiments, the refractive indices of  experimental tank and jet fluid were matched as explained in the \cite{Daviero01}. In the present work, the density and the refractive indices were measured using a densitometer and a refractometer (make $Anton Paar$). 
\par
A dual-head Nd:YAG pulse laser (532 nm, maximum intensity 145 mJ/pulse) is used for both PIV illumination and PLIF excitation. Through PIV optics, the laser beam is expanded into a 1 mm thick laser sheet illuminating the sample area in the x-z plane along the center line of the tank. The jet fluid and ambient medium are uniformly seeded with polyamide tracer particles (median diameter 55 $\mu$m, specific gravity 1.1) for PIV measurement. And for the PLIF measurement, we uniformly mixed the Rhodamine 6G dye in the ambient fluid(in the double buckets) and then the fluid was allowed to stratify in the tank (T2).  To implement the simultaneous PIV/PLIF measurement, the camera lens, PIV filter, PLIF filter, and two cameras, are mounted in an optical housing shown in figure 1. The PIV filter (bandpass, 525nm) blocks most of the fluorescence and passes scattered light from PIV seeding particles. The PLIF filter (high pass in wavelength with cut-off 550 nm) blocks the scattered light and only passes the fluorescence signal. The time delay between the two pulses was set in millisecond. The optical mirror was adjusted such that it illuminates the jet center. An image acquisition and laser control system synchronized the measurements with a sampling rate of 10 Hz. A two-step processing is applied: 64 x 64 pixels interrogation window and 50\% overlap for the first step, and 32 x 32 pixels interrogation window and 50\% overlap for the second step.

\subsection{Image to density field conversion}
The raw images obtained from PLIF camera were dewarped and then processed to get the density field from the below relation (1). In our experiments, we observed that intensity of the background image \& density was found behave approx as a linear function. Based on the linear function, the following equation was used to convert intensity to density. The laser light variation was considered while doing this transformation. The final form of the density formula is
\begin{equation}
\rho =\rho _{j}-\frac{I}{I_{1}(z,t))}\left [ \rho _{j}- \rho_{(z,t)}\right]
\end{equation}
where $\rho _{j}$  is the density of the jet fluid. $I$ is the intensity of the evolving plume and $I_{1}(z,t)$ is the intensity of the background medium, which also takes care of the laser intensity absorption factor in the medium. $\rho_{(z,t)}$ is the density of the background image.

\section{Results \& Discussion}
Using the simultaneous PIV-PLIF technique, various plume parameters were measured and the observations are discussed below.

\subsection{Evolution of plume in the stratified environment}
The mean velocity and density field of a plume evolving in a linearly stratified environment is shown in the figure 2. From the density plot, it is seen that initially the plume fluid has low density and as it moves upwards it entrains with the ambient thereby becoming denser. Similarly, jet velocity decreased as it moves upwards owing to entrainment. In the present experiments, we didn't observe the plume trapping, since the region of the interest in the present study was limited to 19.5 $cm$ from the jet source. The plume trapping was found to occur above the region of interest, because of low stratification strength.

\begin{figure}[h]
	\begin{tabular}{ll}
		\includegraphics[scale=0.4]{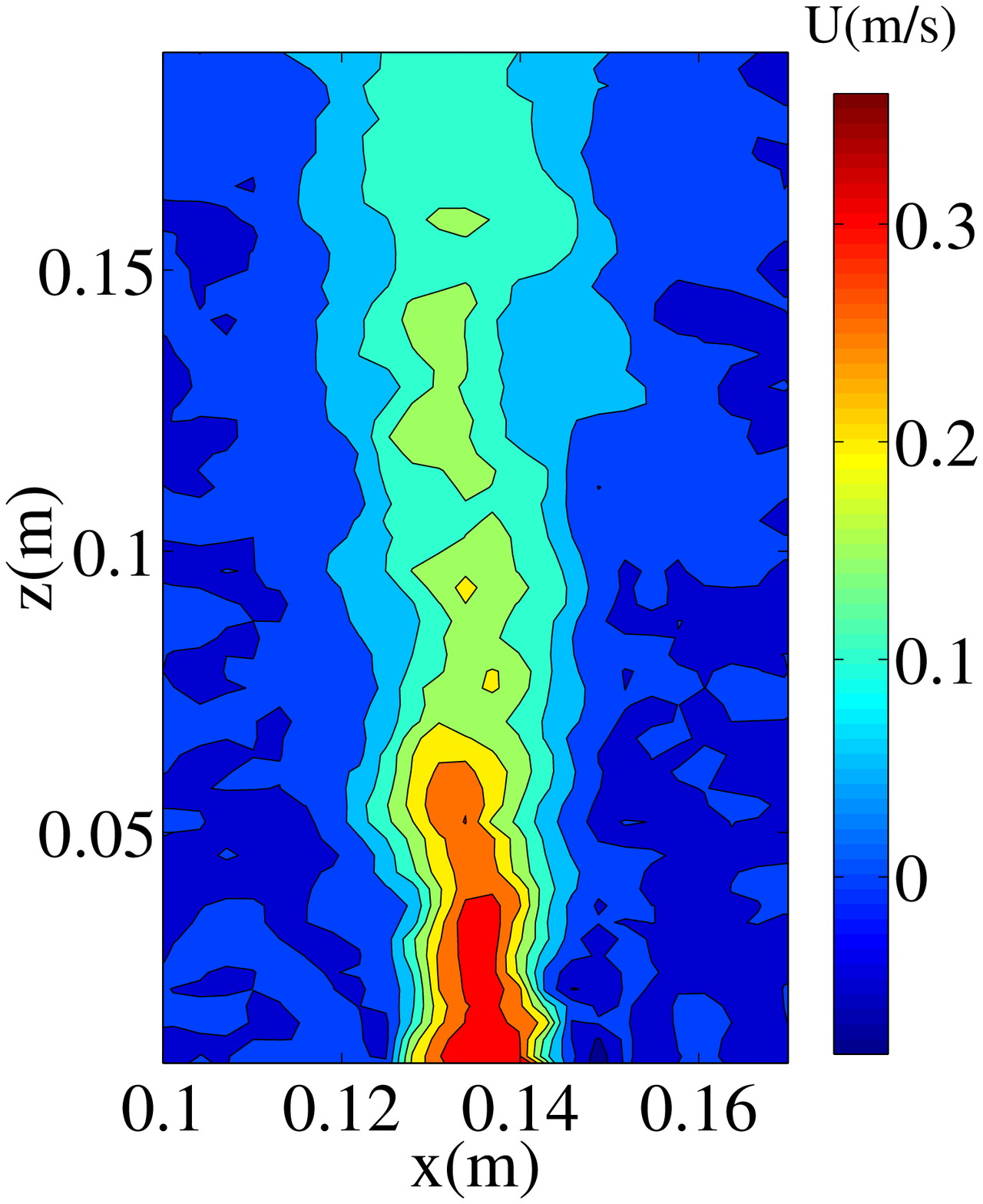}
		&
		\includegraphics[scale=0.61]{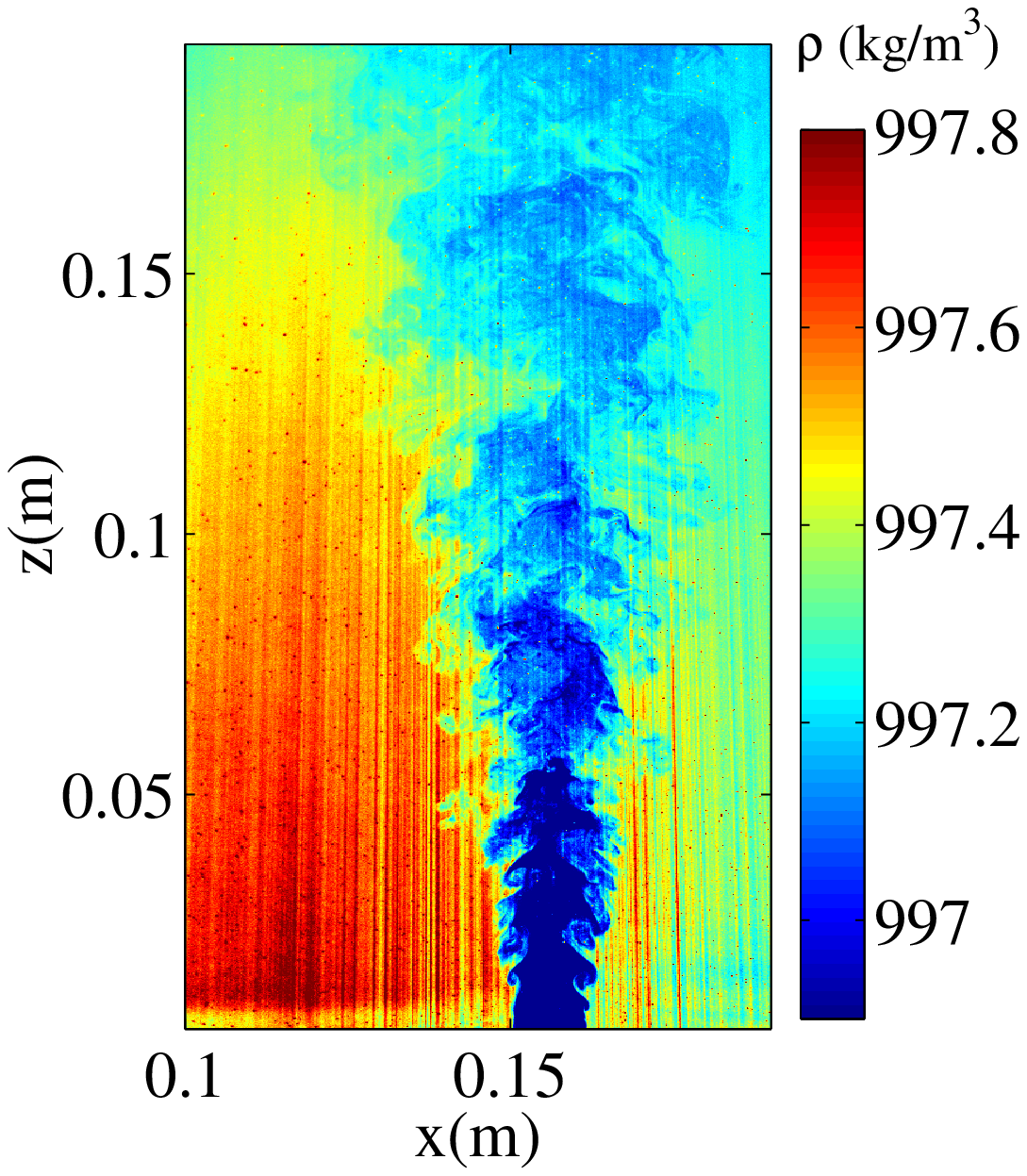}
		
	\end{tabular}
	\caption{Mean Velocity and density contours during quasi-steady plume evolution}
	\label{Fig:Race}
\end{figure}

\subsection{Centerline \& Axial velocity}

The centerline velocity, $U_{c}$, is plotted as a function of axial location, $z$, where the
vertical scale is normalized with respect to the mean velocity $U_{o}$, and the horizontal scale is normalized in relation to $D$ as shown in the  figure 3(a). We observed that the centerline velocity decays linearly proportional as expected for plumes with initial buoyancy flux. 

The axial velocity, $U$, was plotted for the different downstream location is shown in figure 3(b). The mean velocity profile had a symmetric Gaussian profile across its span for the different downstream location. The axial velocity decreases as the plume evolves in the downstream locations, which shows that the momentum of the forced plume is reducing.

\begin{figure}[h]
	\begin{subfigure}[b]{0.5\textwidth}
		\includegraphics[width=\textwidth]{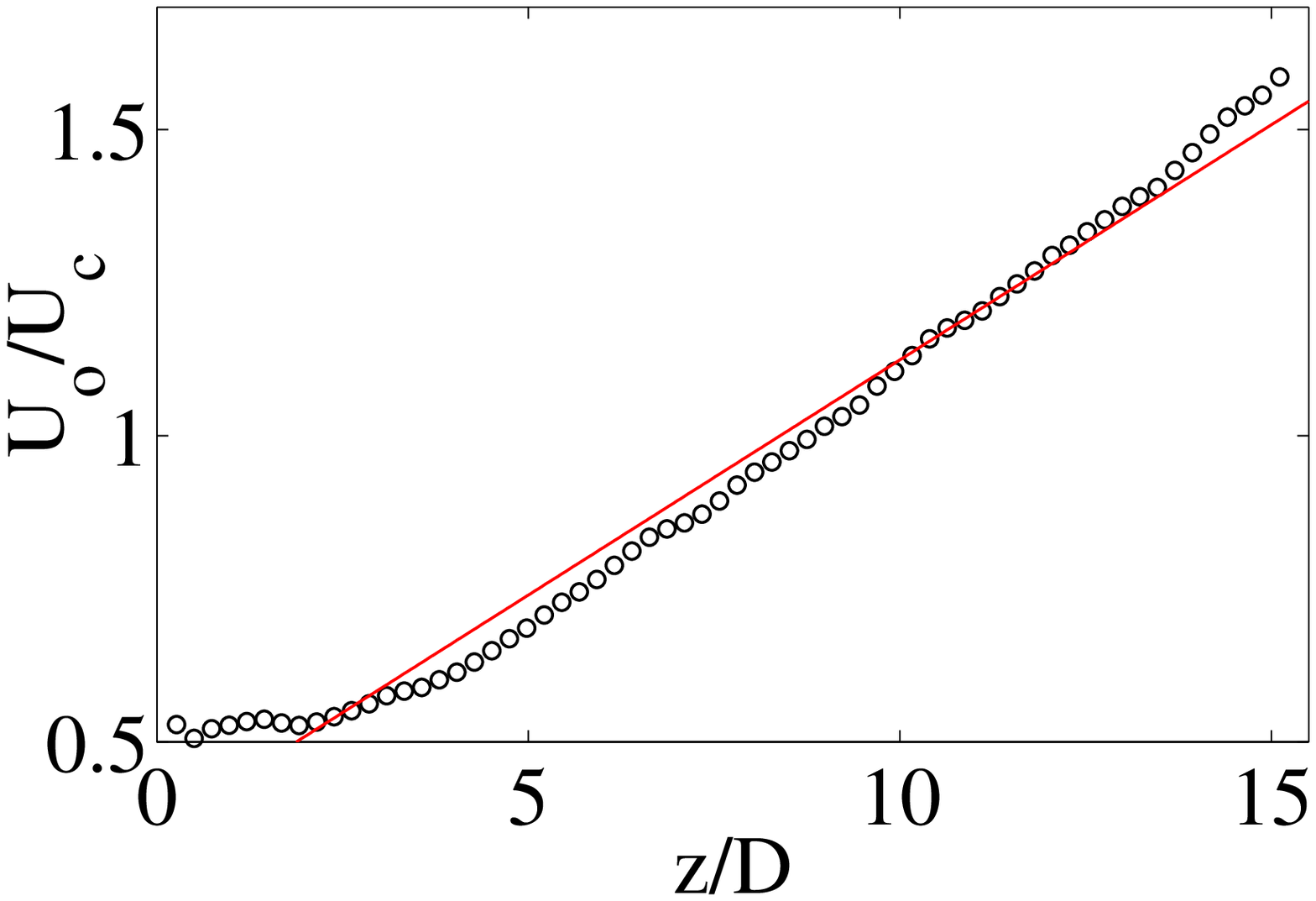}
		\caption{}
		\label{fig:f1}
	\end{subfigure}
	\hfill
	\begin{subfigure}[b]{0.54\textwidth}
		\includegraphics[width=\textwidth]{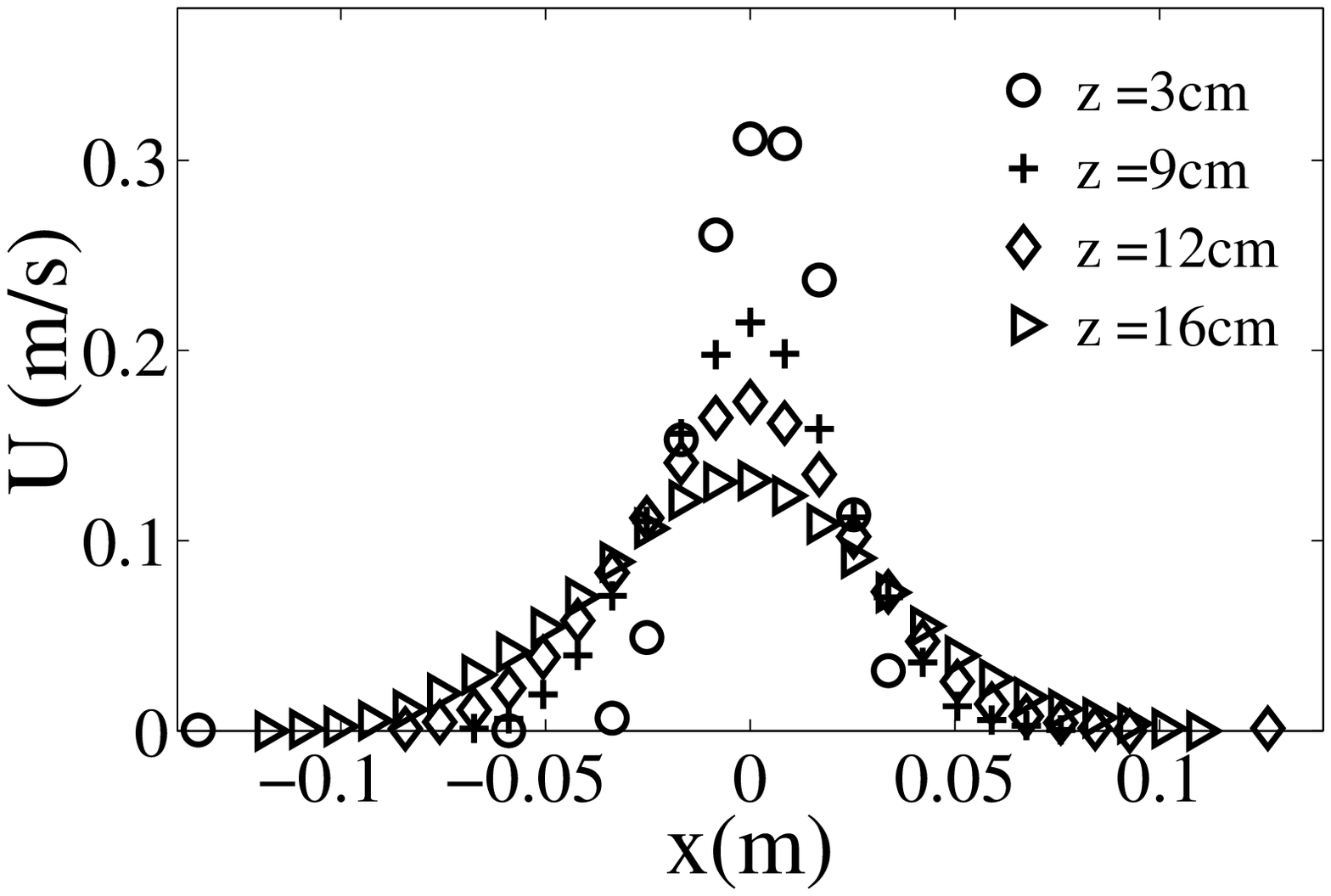}
		\caption{}
		\label{fig:f2}
	\end{subfigure}
\caption{(a): Normalized centerline velocity variation, (-(red) in the plot represent the linear curve fit). (b): Axial velocity variation along the different downstream direction}
\end{figure}

\subsection{Turbulent kinetic energy (K)}
Turbulent kinetic energy, $K$, is one of the most important statistics in stratified flows, which shows the turbulence distribution in the flow. For 2-D flows, this parameter is given as $K=\frac{1}{2}\left ( (\bar{u^{'}})^{2} +(\bar{v^{'}})^2\right )$. We studied the turbulent kinetic energy of the plume at three different downstream locations z/D = 5, 8 \& 12. The normalized turbulent kinetic energy was plotted with the normalized radial coordinate is shown in the figure 4. It was observed that in the case of higher z/D, i.e, near the jet region, the turbulence kinetic energy was more and it gradually decreases with increasing z/D values. Such a behavior is expected since the buoyant plume is losing momentum due to entrainment. The profile of $K$ shows broadening which is attributed to the jet expansion as its moves upwards.

\begin{figure}
\begin{center}
\includegraphics[scale=0.39,trim=10mm 0mm 0mm 0mm]{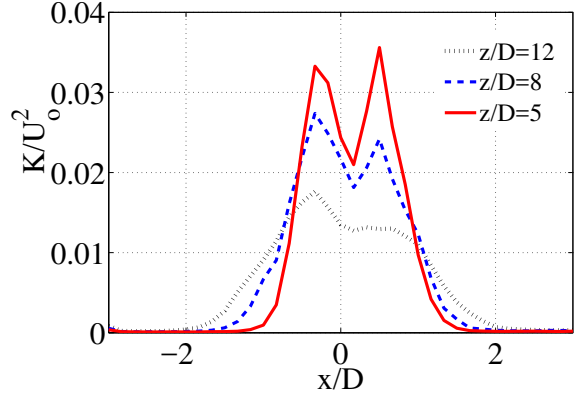}
\caption{Variation of turbulent kinetic energy for the different downstream direction }
\label{fig:1}
\end{center}
\end{figure}

\subsection{Reynolds stresses \& buoyancy fluxes}
The plots of $<u^{'}v^{'}>$ and $<\rho^{'}u^{'}>$  varying along the downstream location are plotted and are shown in the figure 5(a) \& (b). It was observed that the Reynolds stress components decrease with increase in the downstream direction, a trend attributed to the reduction in the magnitude of fluctuating components of velocity due to entrainment. The distribution of $<\frac{u^{'}v^{'}}{U_{o}^{2}} >$  shows a slight off-centre peak that is due to the production of turbulence energy by Reynolds stress working against the mean shear. The profiles obtained for the Reynolds stress are in good agreement to those considered the benchmark literature for turbulent buoyant jets.

The normalized buoyancy flux ${\frac{<\rho^{'}u^{'}>}{\rho _{b}U_{o}}}$ was plotted with normalized  radial location as shown in the figure 5. Where $\rho_{b}$ is bottom density of the stratified medium. The buoyancy flux plot shows the region of stable and unstable fluid motions, which is a feature of stratified flow. The flux value decreases away from the source due to the entrainment of plume and ambient fluid. Further, the flux value $<\rho^{'}u^{'}>$ is order of magnitude lower than $<u^{'}v^{'}>$ indicating momentum dominated flow. Further investigation is in progress to better comprehend the results obtained.  

\begin{figure}[h]
	\begin{subfigure}[b]{0.5\textwidth}
		\includegraphics[width=\textwidth]{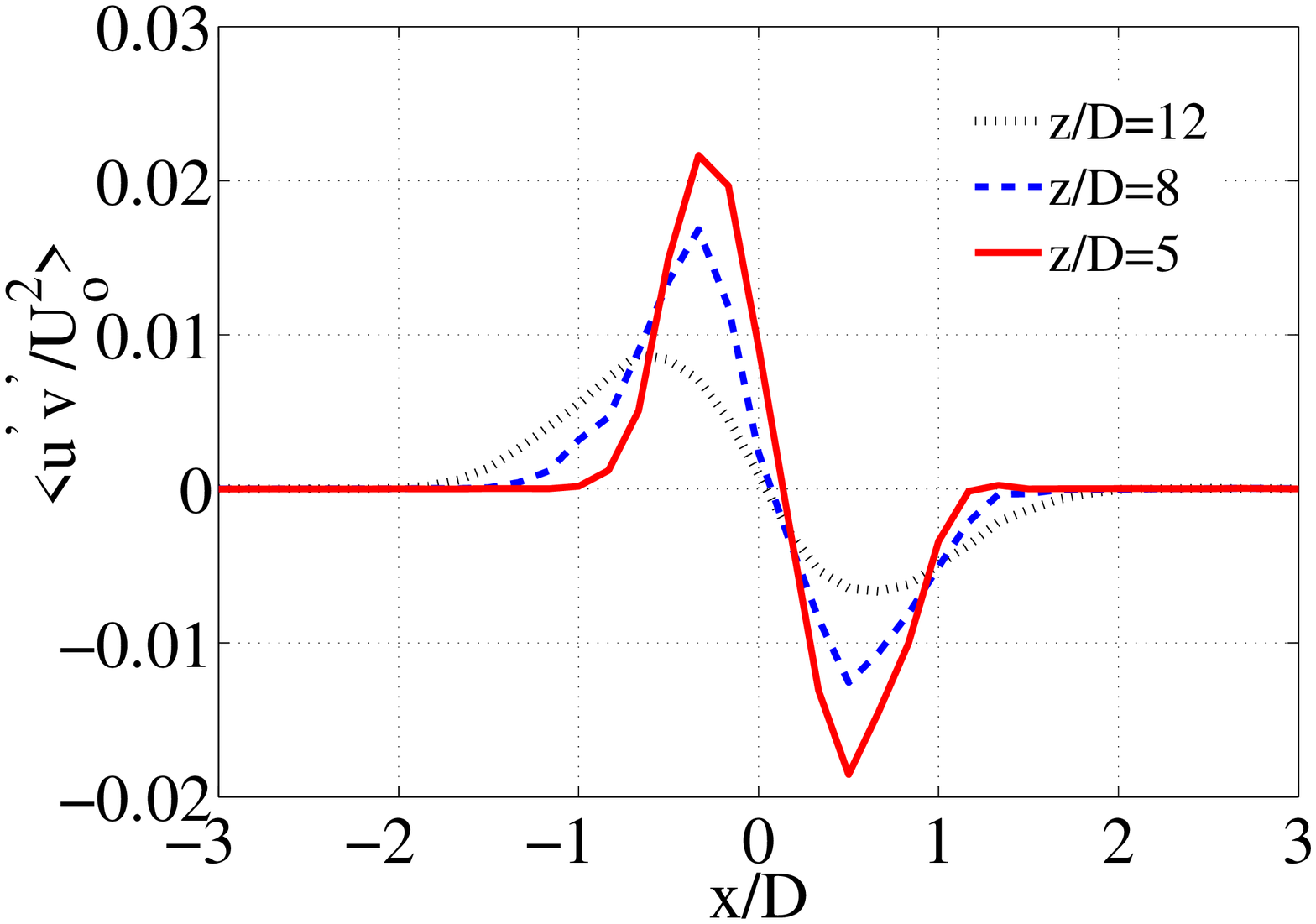}
		\caption{}
		\label{fig:f1}
	\end{subfigure}
	\hfill
	\begin{subfigure}[b]{0.52\textwidth}
		\includegraphics[width=\textwidth]{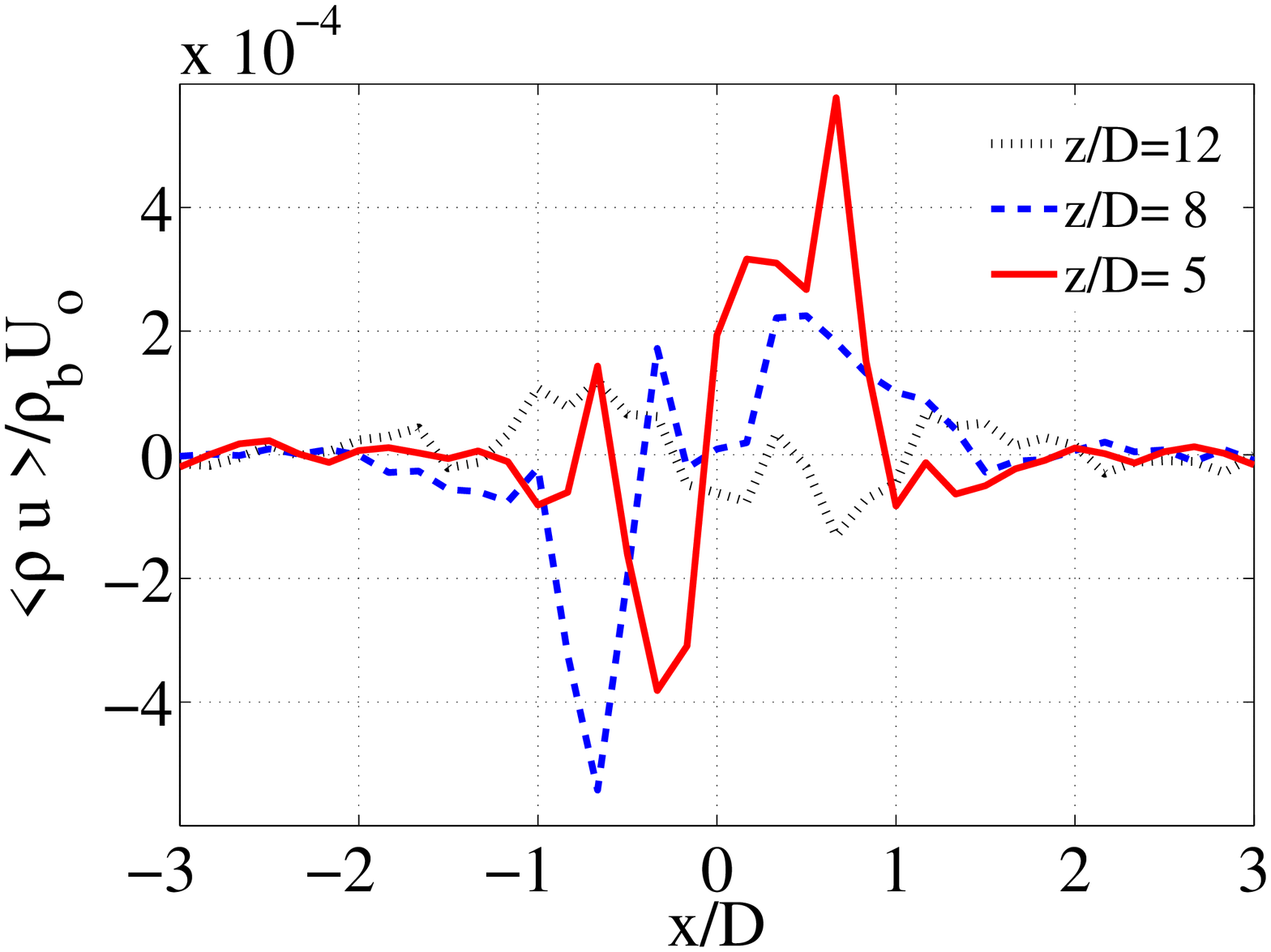}
		\caption{}
		\label{fig:f2}
	\end{subfigure}
	\caption{(a): Variation of $<u^{'}v^{'}>$ for the different downstream location. (b): Variation of  $<\rho^{'}u^{'}>$ for the different downstream location.}
\end{figure}

\section{Conclusion}

The dynamics of a buoyant jet in a linearly stratified environment was studied for the low stratification strength of with density difference 0.5\%, which yields  a value of $N$=0.15 $\mbox{s}^{-1}$. The flow evolution was studied using the simultaneous PIV/PLIF measurement technique, which enables capturing velocity and density fields of an evolving plume. The centerline velocity decays linearly as the plumes evolves in the downstream direction. The axial velocity profile was found to be Gaussian and the value decreased as plume moves away from the source. Turbulent kinetic energy was found to be higher near the jet source, indicating the vigorous mixing, and the value decreases, as the plume moves downstream, owing to entrainment with the ambient fluid. Similar behavior was observed for the Reynolds shear stress component. The buoyancy flux showed stable and unstable nature of the plume, thereby presents insight to the plume mixing in a stratified medium. Further experimental work are in progress to understand the detailed plume behavior in the far field and also to understand the plume dynamics for the high stratification strength. 

\acknowledgements
The authors acknowledge funding from Department of Science $\&$ Technology, and Ministry of Earth Sciences for the present research work. We further acknowledge the help rendered by Mr. Rajesh Chauhan \& Mr. Shashank Benadikar in conducting the experiments.

\references
\def\refname{}
\vspace{-0.9in}
\bibliographystyle{apalike}
\bibliography{references}   

\end{document}